\documentclass[letterpaper,twocolumn,english,floatfix,showpacs,amsmath,amssymb,reprint,nofootinbib]{revtex4-1}
\usepackage[T1]{fontenc}
\usepackage[latin9]{inputenc}
\usepackage{color}
\usepackage{babel}
\usepackage{amsmath}
\usepackage{amssymb}
\usepackage{graphicx}
\usepackage[unicode=true,pdfusetitle,
 bookmarks=true,bookmarksnumbered=false,bookmarksopen=false,
 breaklinks=true,pdfborder={0 0 0},backref=false,colorlinks=true]
 {hyperref}
\hypersetup{
 linkcolor=blue,citecolor=blue,urlcolor=blue}
\usepackage{breakurl}

\makeatletter

\usepackage{babel}

\IfFileExists{lmodern.sty}{\usepackage{lmodern}}{}

\makeatother

\begin{document}

\title{Entanglement properties of correlated random spin chains and similarities
with conformally invariant systems }

\author{João C. Getelina}

\affiliation{Instituto de Física de São Carlos, Universidade de São Paulo, C.P.
369, São Carlos, SP 13560-970, Brazil}

\author{Francisco C. Alcaraz}

\affiliation{Instituto de Física de São Carlos, Universidade de São Paulo, C.P.
369, São Carlos, SP 13560-970, Brazil}

\author{José A. Hoyos}

\affiliation{Instituto de Física de São Carlos, Universidade de São Paulo, C.P.
369, São Carlos, SP 13560-970, Brazil}
\begin{abstract}
We study the Rényi entanglement entropy and the Shannon mutual information
for a class of spin-1/2 quantum critical XXZ chains with random coupling
constants which are partially correlated. In the XX case, distinctly
from the usual uncorrelated random case where the system is governed
by an infinite-disorder fixed point, the correlated-disorder chain
is governed by finite-disorder fixed points. Surprisingly, we verify
that, although the system is not conformally invariant, the leading
behavior of the Rényi entanglement entropies are similar to those
of the clean (no randomness) conformally invariant system. In addition,
we compute the Shannon mutual information among subsystems of our
correlated-disorder quantum chain and verify the same leading behavior
as the $n=2$ Rényi entanglement entropy. This result extends a recent
conjecture concerning the same universal behavior of these quantities
for conformally invariant quantum chains. For the generic spin-1/2
quantum critical XXZ case, the true asymptotic regime is identical
to that in the uncorrelated disorder case. However, these finite-disorder
fixed points govern the low-energy physics up to a very long crossover
length scale and the same results as in the XX case apply. Our results
are based on exact numerical calculations and on a numerical strong-disorder
renormalization group. 
\end{abstract}


\maketitle

\section{Introduction}

The entanglement entropy has been proven as an important tool to study
and classify statistical mechanics and condensed matter systems. For
pure state systems, the most used measure of the entanglement between
two complementary regions ${\cal A}$ and ${\cal B}$ is the so-called
Rényi entanglement entropy 
\begin{equation}
S_{n}\left({\cal A},{\cal B}\right)=\frac{1}{1-n}\ln{\rm Tr}\rho_{{\cal A}}^{n},\label{eq:S-Renyi-def}
\end{equation}
with $\rho_{{\cal A}}={\rm Tr}_{{\cal B}}\rho$ being the reduced
density matrix of subsystem ${\cal A}$. For $n\rightarrow1$, it
recovers the von Neumann entanglement entropy $S_{1}=S_{{\rm vN}}=-{\rm Tr}\rho_{{\cal A}}\ln\rho_{{\cal A}}$.

In the arena of one-dimensional conformally invariant quantum systems,
such measure allows us to identify the various distinct universality
classes of critical behavior\ \cite{holzhey-larsen-wilczek,calabrese-cardy-jstatmech04}.
It is now understood that the ground state entanglement entropy reads\ \cite{calabrese-etal-prl10,xavier-alcaraz-prb11}
\begin{equation}
S_{n}\left(x,L\right)=\frac{c}{6}\left(1+\frac{1}{n}\right)\ln f_{L}^{{\rm CFT}}(x)+\kappa_{1}+\frac{\left(-1\right)^{x}\kappa_{2}}{\left(f_{L}(x)\right)^{\phi/n}}+\dots,\label{eq:CFT-entanglement}
\end{equation}
where $c$ is the universal central charge of the underlying conformal
field theory (CFT), $L$ is the system size (here, we consider periodic
boundary conditions), $1\ll x\leq L/2$ is the size of the subsystem
${\cal A}$ , $f_{L}^{{\rm CFT}}\left(x\right)=\frac{L}{\pi}\sin\left(\frac{\pi x}{L}\right)$
is the scaling function (which equals the chord length), $\kappa_{1,2}$
are nonuniversal constants, and $\phi$ is related to the scaling
dimension of the energy operator. For $n=1$ or for systems without
a Fermi surface, $\kappa_{2}=0$.

However, there is no simple way of accessing the Rényi entanglement
entropy directly from experimental measurements. It is then desirable
to study other information measures that can be, in principle, directly
observed in experiments, and, like $S_{n}$, enable us to determine
the critical universality class of the system. Recently, the Shannon
mutual information was proposed as one such information measure\ \cite{alcaraz-rajabpour-prl13,alcaraz-rajabpour-prb14}.
It is defined as 
\begin{equation}
I\left({\cal A},{\cal B}\right)={\rm Sh}\left({\cal A}\right)+{\rm Sh}\left({\cal B}\right)-{\rm Sh}\left({\cal A}\cup{\cal B}\right),\label{eq:I-Shannon-def}
\end{equation}
where ${\rm Sh}\left({\cal A}\right)=-\sum_{m}p_{m}\ln p_{m}$ is
the usual Shannon entropy, with $p_{m}$ being the probability of
finding the subsystem ${\cal A}$ in a configuration $m$ obtained
from the normalized wave function $\left|\psi_{{\cal A}\cup{\cal B}}\right\rangle =\sum_{m,n}c_{m,n}\left|\phi_{{\cal A}}\right\rangle _{m}\otimes\left|\phi_{{\cal B}}\right\rangle _{n}$,
namely $p_{m}=\sum_{n}\left|c_{m,n}\right|^{2}$. In general the Shannon
entropy is a basis-dependent quantity. However, it was conjectured
that $I({\cal A},{\cal B})$, in some special bases, shares the same
leading asymptotic behavior as $S_{2}$ given by Eq.\ (\ref{eq:CFT-entanglement}).
This result was verified numerically for several quantum chains in
distinct universality classes.\ %
\footnote{In the case of the quantum Ising chain, criticisms can be found in
Ref.\ \onlinecite{stephan-prb14}.%
}

The entanglement properties of quantum critical random (quenched disordered)
systems are much less understood. All our knowledge is related to
random spin chains governed by exotic infinite-randomness fixed point
(IRFP)\ \cite{fisher94-xxz,fisher95}. Since the ground state of
these fixed points can be understood as a collection of nearly independent
spin clusters arranged in a fractal fashion, it can be easily shown
that $S_{n}\sim\gamma\ln f_{L}^{{\rm IRFP}}\left(x\right)$, with
$\gamma=\ln2$ being a universal $n$-independent constant\ \cite{refael-moore-prl04,laflorencie-entanglement,hoyos-rigolin,santachiara-06,hoyosvieiralaflorenciemiranda,bonesteel-yang-prl07,fidkowski-etal-prb08,fagotti-calabrese-moore-prb11,tran-bonesteel-prb11,pouranvari-yang-prb13,ramirez-etal-jsm14}
in contrast to the clean system {[}see Eq.\ (\ref{eq:CFT-entanglement}){]}.
This understanding stems exclusively from the analytical tool of the
strong-disorder renormalization-group (SDRG) method\ \cite{MDH-PRL,MDH-PRB,bhatt-lee}
which can be used to compute $S_{n}$\ \cite{refael-moore-prl04,hoyosvieiralaflorenciemiranda}.
More recently, the scaling function $f_{L}^{{\rm IRFP}}(x)$ have
been studied and shown to differ from the chord length\ \cite{fagotti-calabrese-moore-prb11},
namely 
\begin{equation}
f_{L}^{{\rm IRFP}}\left(x\right)\approx\frac{L}{\pi}\left(\sin\left(\frac{\pi x}{L}\right)-0.153\sin^{3}\left(\frac{\pi x}{L}\right)\right).\label{eq:f-IRFP}
\end{equation}

The entanglement properties of quantum critical random systems governed
by conventional finite-disorder fixed points are much less known.
A possible reason is due to the lack of simple analytical and numerical
tools for handling them. Examples of such systems are frustrated spin
ladders or quantum spin chains with random ferro- and antiferromagnetic
interactions which are of great interest from the experimental\ \cite{Nguyen-etal-science96,irons-etal-prb00,koteswararao-etal-jp10}
and theoretical\ \cite{westerberg-PRB-FM-AF,yusuf-zigzag,hoyos-ladders,lavarelo-etal-prb13}
points of view. However, even though a strong-disorder renormalization-group
description of these systems is possible, unlike the infinite-disorder
case, the entanglement entropy $S_{n}$ cannot be computed in a simple
way. 

Apparently, the only known system governed by a finite-disorder fixed
point whose entanglement properties were studied, namely the von Neumann
entropy $S_{1}$, is the quantum Ising chain with correlated disorder\ \cite{binosi-entanglement-prb}.
In this special case it is understood that disorder correlation prevents
the generation of random mass, yielding thus a perturbatively irrelevant
disorder\ \cite{hoyos-etal-epl11}. Further increase of the disorder
strength, beyond the perturbative limit, drives the system towards
a line of finite-disorder critical fixed points. Interestingly, along
this line, the ground state entanglement entropy increases.

In this paper we report an extensive study of Rényi entanglement entropy
$S_{n}(x,L)$ and of the Shannon mutual information $I(x,L)$ for
the critical XX spin chain (which is equivalent to a doubled quantum
Ising chain at criticality) with correlated disorder. We show that
$S_{n}$ exhibts the same leading finite-size scaling function $f_{L}^{{\rm CFT}}$
as in (\ref{eq:CFT-entanglement}) with a different prefactor (\emph{effective
central charge}), in contrast to the infinite-randomness case (\ref{eq:f-IRFP}).
In addition, our results indicate that the Shannon mutual information
$I(x,L)$ shares the same asymptotic behavior as $S_{2}$, as conjectured
for the conformally invariant clean system. Furthermore, we show that
these features are not exclusive particularities of the noninteracting
XX model but are also present in the critical phase of the XXZ spin-1/2
chain below a relatively large crossover length.

The remainder of this article is organized as follows. In Sec.\ \ref{sec:Model}
we introduce the model and the numerical methods used. In Sec.\ \ref{sec:Dynamical-exponent}
we compute the dynamical critical exponent, that indicates the presence
of a line of finite-disorder fixed points. The entanglement measures
$S_{n}$ and $I$ are computed in Sec.\ \ref{sec:Entanglement},
and their dependence on the disorder strength and anisotropy is analyzed
in detail. Our SDRG study is reported in Sec.\ \ref{sec:SDRG}, where
we derive the renormalization-group procedure and explain its numerical
implementation for very large chains. Conclusions and final remarks
are given in Sec.\ \ref{sec:Conclusion}. For completeness, in Appendices
\ref{sec:App} and \ref{sec:3-spins} we compute the entanglement
measures for random singlet states, and our SDRG decimation rules,
respectively.

\section{The model and methods\label{sec:Model}}

We are interested in the ground state entanglement properties of the
XXZ spin-1/2 chain given by the Hamiltonian 
\begin{equation}
H=\sum_{i=1}^{L}J_{i}\left(S_{i}^{x}S_{i+1}^{x}+S_{i}^{z}S_{i+1}^{z}+\Delta S_{i}^{z}S_{i+1}^{z}\right),\label{eq:Hamiltonian}
\end{equation}
where $\mathbf{S}_{i}$ are spin-1/2 operators. The lattice size is
$L$ (even) and we consider periodic boundary conditions $\mathbf{S}_{L+1}=\mathbf{S}_{1}$.
The anisotropy $\Delta$ is site independent and we constrain our
analysis to $-1<\Delta\leq1$, a region where the clean model is critical
and conformally invariant. Quenched disorder is implemented via the
coupling constants $J_{i}$ which are drawn from the distributions
\begin{equation}
P(J)=\frac{1}{D\Omega_{I}}\left(\frac{\Omega_{I}}{J}\right)^{1-1/D},\mbox{ with }0<J<\Omega_{I}.\label{eq:P(J)}
\end{equation}
Here, the disorder strength is parametrized by $D$ and $\Omega_{I}$
sets the energy scale. Correlation among the disorder variables is
implemented as follows. All odd-numbered coupling constants are independent
random variables drawn from Eq.\ (\ref{eq:P(J)}). The even-numbered
couplings, however, are equal to the odd one to their left, i.e.,
$J_{2i}=J_{2i-1}$. In this way, the sequence of couplings in our
random correlated spin chain reads $J_{1},J_{1},J_{3},J_{3},\dots J_{L-1},J_{L-1}$.

As shown in Ref.\ \onlinecite{hoyos-etal-epl11}, for weak disorder
$D\leq D_{c}\approx0.3$ and for the XX model ($\Delta=0$), the system
is governed by the clean critical point. For larger disorder values
$D>D_{c}$, a line of finite-disorder fixed points is accessed. This
contrasts with the uncorrelated case ($J_{i}$ are independent random
variables $\forall i$), where disorder is perturbatively relevant
and drives the system to a random singlet state described by a universal
renormalization-group fixed point of infinite-randomness nature\ \cite{fisher94-xxz}.

The essence of this difference is the absence of a random mass in
the former case. For $\Delta=0$, the energy mass gap is given by
a function of the ratio between the products of even- and odd-numbered
couplings\ \cite{pfeuty-pla79}: $m\sim\sum_{i=1}^{L/2}\ln(J_{{\rm 2i}}/J_{{\rm 2i+1}})$.
Therefore, it is reasonable to define $m_{x}=\overline{\ln J_{{\rm odd}}}-\overline{\ln J_{{\rm even}}}$
(where $\overline{\cdots}$ means the average over a coarse-grained
region $x$) as a measure of the local distance from criticality,
i.e., the local mass. This is the argument used in Ref.\ \onlinecite{hoyos-etal-epl11}
to ensure that the effective action of the system does not contain
a random mass term. In the generic case $\Delta\neq0$, the dependence
of mass gap $m$ on the random couplings $J_{i}$ is not known. Presumably,
it depends on the anisotropy $\Delta$ and will not be given just
by a simple ratio among products of $J_{i}$. We then expect that
the disorder couples to the mass term, and therefore the true ground
state should be a random singlet just like in the uncorrelated disorder
case. As we will see below, however, the random mass term seems to
be quite small, and its effects are only seen at very low energy scales
(very large chains). Hence, the physics of the quenched disordered
quantum XXZ spin-1/2 chain without random mass can be studied in the
preasymptotic regime of smaller chains.

The model in Eq.\ (\ref{eq:Hamiltonian}) is studied via two complementary
methods: exact diagonalization (see Secs.\ \ref{sec:Dynamical-exponent}
and \ref{sec:Entanglement}) and strong-disorder renormalization group
(see Sec.\ \ref{sec:SDRG}).

In the case $\Delta=0$, a map to free fermions enables us to exactly
diagonalize the system for quite large sizes. For $\Delta\neq0$,
we use the power method to evaluate the low-lying states up to system
sizes $L=22$. Since the fixed points are of finite-disorder nature,
the statistical fluctuations are much weaker as compared to those
of the infinite-disorder fixed point. Then, disorder averaging over
$N=10^{3}$ samples is sufficient for a reasonable accuracy in most
of the cases.

When we use the SDRG method, the anisotropy values $\Delta=0$ and
$\Delta\neq0$ can be treated on an equal footing and chains of sizes
$L=10^{6}$ and $10^{7}$ can be easily reached.

\section{The dynamical critical exponent\label{sec:Dynamical-exponent}}

\begin{figure}[t]
\begin{centering}
\includegraphics[clip,width=0.7\columnwidth]{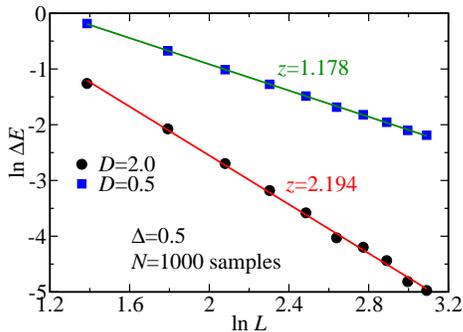} 
\par\end{centering}

\caption{(Color online) Finite-size gap $\Delta E$ for various system sizes
$L=4,6,8,\dots,22$, for the cases of disorder strength $D=0.5$ (blue
squares) and $2.0$ (black circles) and anisotropy $\Delta=0.5$.
The lines are the best linear fits, and the corresponding estimated
values of $z$ are also shown.\label{fig:Finite-size-gap}}
\end{figure}

In this section we compute the dynamical critical exponent $z$ obtained
from the leading behavior of the finite-size energy gap 
\begin{equation}
\Delta E\sim L^{-z},\label{eq:z}
\end{equation}
which is plotted in Fig.\ \ref{fig:Finite-size-gap} for the anisotropy
$\Delta=1/2$ and the disorder strengths $D=0.5$ and $2$. The continuous
lines in the figure are the best fits for $z$.

It is important to remark that there is no crossover length $\xi$,
in contrast with the uncorrelated disorder case. In the latter case,
the true infinite-randomness fixed point is reached only when $L$
exceeds a disorder-dependent crossover length $\xi$. Below this length,
the clean fixed point governs the system. Differently from Fig.\ \ref{fig:Finite-size-gap},
we would get a dynamical critical exponent $z=1$ for $L\ll\xi$ that
crosses over to $z\rightarrow\infty$ for $L\gg\xi$. This crossover
phenomena is observed in many quantities for the case of \emph{uncorrelated}
disorder\ \cite{laflorencie-correlacao-PRB,hoyosvieiralaflorenciemiranda}.
However, in the case of \emph{correlated} disorder, such crossover
behavior is absent.

\begin{figure}[b]
\begin{centering}
\includegraphics[clip,width=0.7\columnwidth]{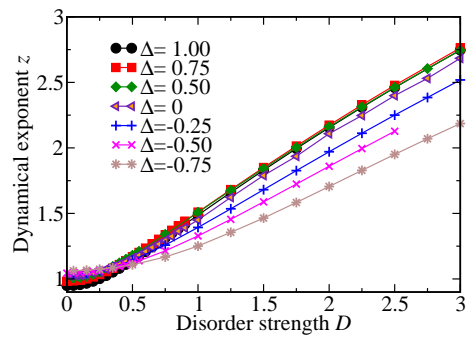} 
\par\end{centering}

\caption{(Color online) Dynamical exponent $z$ as a function of the disorder
strength $D$ for several anisotropy parameter $\Delta$. The results
were obtained from the fit of the finite-size gap as in Fig.\ \ref{fig:Finite-size-gap}.
The estimated error bars are of the order of the symbol sizes and
the lines are guides to the eyes.\label{fig:dynamical-exponent}}
\end{figure}

The critical exponent $z$ is plotted in Fig.\ \ref{fig:dynamical-exponent}
for various disorder strengths $D$ and anisotropies $\Delta$. As
can be noticed, for $D<D_{c}\left(\Delta\right)$ the exponent $z$
is approximately given by the clean theory value $z_{{\rm clean}}=1$.
We also verified that other critical exponents for $D<D_{c}$ are
equal to the clean ones, i.e., for $D<D_{c}$ the system is in the
same universality class as the clean model. Interestingly, $D_{c}\left(\Delta\right)\approx0.3$
seems to be $\Delta$ independent. We believe that this is indeed
the case.

\begin{figure}[t]
\begin{centering}
\includegraphics[clip,width=0.9\columnwidth]{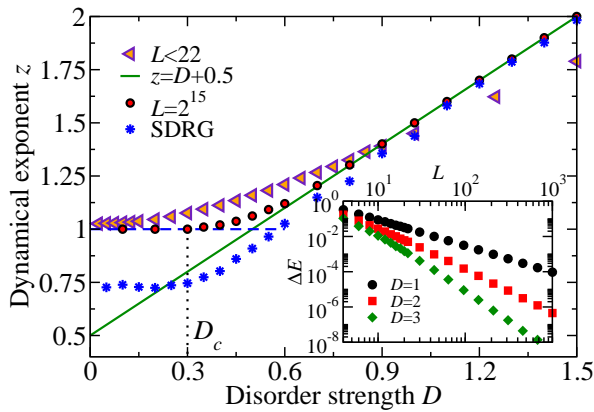} 
\par\end{centering}

\caption{(Color online) Comparison of the dynamical critical exponent for the
$\Delta=0$ case computed by exact diagonalizations of small chains
($L<22$), of very large chains ($L=2^{15}$), and via the SDRG method
($L=10^{6}$). Inset: The finite-size gap for different system sizes
where the corrections to scaling are clear. Error bars are about the
same size as symbols. \label{fig:Finite-size-effects} }
\end{figure}

As can be guessed from Figs.\ \ref{fig:Finite-size-gap} and \ref{fig:dynamical-exponent},
due to the available small system sizes, finite-size corrections to
the scaling of Eq.\ (\ref{eq:z}) are not negligible. We expect $\Delta E\sim L^{-z}\left(1+AL^{-\omega}\right)$,
where the constants $A$ and the exponent $\omega$ (which were not
calculated) take care of the leading correction to scaling. In order
to better see them, we plot in Fig.\ \ref{fig:Finite-size-effects}
the exponent $z$ for the anisotropy $\Delta=0$, since bigger lattice
sizes can be reached. The results for large chains ($L=2^{15}$) are
reproduced from Ref.\ \onlinecite{hoyos-etal-epl11} and the finite-size
effects are negligible. The inset shows the finite-size gaps for many
system sizes up to $L=10^{3}$. For disorder strength $D=2$ and $3$
it becomes clear that corrections to scaling exist for smaller chains.

Notice that, for $\Delta\geq0$, our results strongly suggest that
$z$ is $\Delta$ independent (see Fig.\ \ref{fig:dynamical-exponent}).
For $\Delta<0$, we see a small dependence. However, due to the strong
finite-size effects, we cannot exclude the possibility of $z$ also
being $\Delta$ independent for those negative values.

As we shall see in Secs.\ \ref{sec:Entanglement} and \ref{sec:SDRG},
the entanglement properties and the SDRG analysis also suggest this
$\Delta$ independence.

\section{The Rényi entanglement entropies and the Shannon mutual information\label{sec:Entanglement}}

In this section we numerically compute the Rényi entanglement entropy
$S_{n}\left(x,L\right)$ and the Shannon mutual information $I\left(x,L\right)$
of subsystems of size $x$ of an $L$-site XXZ quantum chain. We restrict
ourselves to the density matrix formed by the ground state eigenfunction.

We verify that for $D<D_{c}$ these quantities show the same asymptotic
behavior as occurs in the clean model {[}see Eq.\ (\ref{eq:CFT-entanglement}){]}.
This is not a surprise since, as discussed in Sec.\ \ref{sec:Dynamical-exponent},
they share the same critical universality class.

\begin{figure}[t]
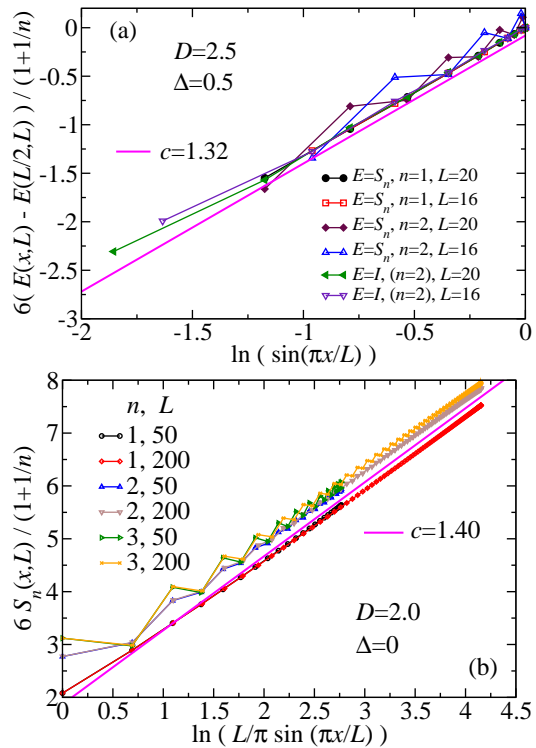

\begin{centering}
\includegraphics[clip,width=0.8\columnwidth]{fig4a}\\
 \includegraphics[clip,width=0.8\columnwidth]{fig4b} 
\par\end{centering}

\caption{(Color online) Entanglement measures for the correlated disordered
XXZ quantum spin chain with (a) anisotropy $\Delta=0.5$ and disorder
$D=2.5$ and for (b) anisotropy $\Delta=0$ and disorder $D=2.0$
for several indices $n$ and lattice sizes $L$. Entanglement is measured
by the Rényi entanglement entropy $S_{n}$ ($n=1,\,2,\mbox{ and }3$)
and by the Shannon mutual information $I$. Data are averaged over
$10^{3}$ disorder realizations and the lines are guides to the eyes.
\label{fig:Entanglement-entropy}}
\end{figure}

Surprisingly, even for $D>D_{c}$ the finite-size scaling behavior
of $S_{n}\left(x,L\right)$ and $I\left(x,L\right)$ are the same
as in the clean system. The difference is only in the prefactor ``effective
central charge.''%
\footnote{Strictly speaking, the central charge is only defined for systems
in which $z=1$. Nevertheless, the effective central charge $c^{{\rm eff}}$,
similarly to what happens in conformally invariant quantum chains,
measures the effective size $\left(\sim x^{c^{{\rm eff}}}\right)$
of the entangled Hilbert space of the reduced subsystem of size $x$.%
} This is illustrated in Fig.\ \ref{fig:Entanglement-entropy}, where
we plot $S_{n}\left(x,L\right)$ for two values of anisotropies and
disorder strengths: $\Delta=1/2$ and $D=2.5$ {[}panel (a){]} and
$\Delta=0$ and $D=2.0$ {[}panel (b){]}. Several values of $n$ and
length sizes $L$ are shown in the same plot. As can be clearly seen,
apart from small oscillations in $S_{n>1}$, all data seems to collapse
in a single universal curve. This means that $S_{n}\left(x,L\right)$
and $I\left(x,L\right)$ share the same asymptotic finite-size scaling
function as conformally invariant quantum chains {[}see Eq.\ (\ref{eq:CFT-entanglement}){]},
although there is no conformal invariance for those random systems.

The straight solid line is the linear fit for $4\leq x\leq L/2$,
from which we extract the effective central charge $c$ values: $1.32$
and $1.40$, for $D=2.5$, $\Delta=0.5$ and $D=2.0$, $\Delta=0$,
respectively.

Interestingly, the Shannon mutual information $I$ in Fig.\ \hyperref[fig:Entanglement-entropy]{\ref{fig:Entanglement-entropy}(a)}
has the same asymptotic behavior of $S_{n=2}$.

The Shannon entropy ${\rm Sh}(x,L)$ and consequently the mutual information
$I(x,L)$ are, in general, basis-dependent quantities in contrast
with the Rényi entanglement entropy which is basis independent. In
Fig.\ \hyperref[fig:Entanglement-entropy]{\ref{fig:Entanglement-entropy}(a)}
we computed $I(x,L)$ in the $S^{z}$-basis. Universal behavior for
$I(x,L)$ in the conformally invariant systems happens only when the
ground state is expressed in the so-called conformal basis that corresponds
in the case of the clean XXZ model to the $S^{z}$ and $S^{x}$ bases\ \cite{alcaraz-rajabpour-prb14}.
Although we did not compute $I(x,L)$ for our correlated random system
using the $S^{x}$ basis, we do not expect any different behavior
in this basis.

It is worth mentioning that $I\left(x,L\right)$ does not display
the oscillatory parity effect of $S_{2}\left(x,L\right)$, which implies
they are indeed different quantities with different subleading terms.

Finally, the fact that we have a straight line even for small subsystem
sizes is in contrast to the uncorrelated case, where there is a crossover
subsystem size $\xi$ separating the behavior of the clean and disordered
systems. This is in agreement with the absence of a crossover length
in the correlated disorder case as seen in Sec.\ \ref{sec:Dynamical-exponent}.

In the remainder of this Section, we further explore two other aspects
of our numerical results: (i) the prefactor $c_{D}^{{\rm eff}}(1+1/n)/6$
and $c_{D}^{{\rm eff}}/4$ of the leading terms of $S_{n}$ and $I$,
respectively, and (ii) the functional dependence of the scaling function
with the chord length $\frac{L}{\pi}\sin\frac{\pi x}{L}$.

\begin{figure}[t]
\begin{centering}
\includegraphics[clip,width=0.8\columnwidth]{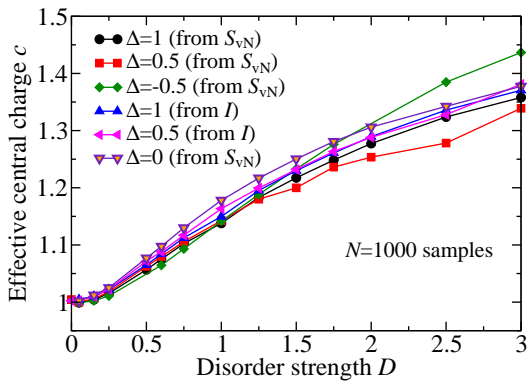} 
\par\end{centering}

\caption{(Color online) The effective central charge as a function of disorder
strength $D$ obtained from the von Neumann entanglement entropy $S_{{\rm vN}}$
and from the Shannon mutual information $I$ for various anisotropies
$\Delta$. For $\Delta\protect\neq0$, these values are effective
ones due to the system finite size. For extremely large systems, these
values crossover to $c_{{\rm IRFP}}^{{\rm eff}}=\ln2$ (see Sec.\ \ref{sec:SDRG}).
\label{fig:C-charge}}
\end{figure}

For spin chains governed by infinite-randomness fixed points, it is
easy to show that the Rényi entanglement entropy $S_{n}\sim\frac{1}{3}c_{{\rm IRFP}}^{{\rm eff}}\ln x$
for any $n$ and $1\ll x\ll L$; i.e., the numerical prefactor $c_{{\rm IRFP}}^{{\rm eff}}$
does not depend on $n$ (see Appendix\ \ref{sec:App}). The fact
that our spin chains with correlated disorder do not follow this prediction
is another indication that the ground state is not a collection of
``random singlets.'' As already discussed, our numerical results
indicate that the leading term has the same dependence on $n$ as
conformally invariant systems. In order to check the dependence of
$c_{D}^{{\rm eff}}$ with $\Delta$ we plot in Fig.\ \ref{fig:C-charge}
$c_{D}^{{\rm eff}}$ as a function of $D$ for various values of $\Delta$
and for the lattice sizes $L=14,\,16,\,18$, and $20$. The data are
averaged $10^{3}$ disorder realizations. Analogously to what happens
with the critical dynamical exponent $z$, the effective central charge
exhibits quite small variations with the anisotropy $\Delta$. Since
we used only relatively small lattice sizes, and we notice the strong
statistical fluctuations (especially for higher $D$), it is then
plausible to expect that $c_{D}^{{\rm eff}}$ does not depend on $\Delta$,
being only a function of $D$.

\begin{figure}[b]
\begin{centering}
\includegraphics[clip,width=1\columnwidth]{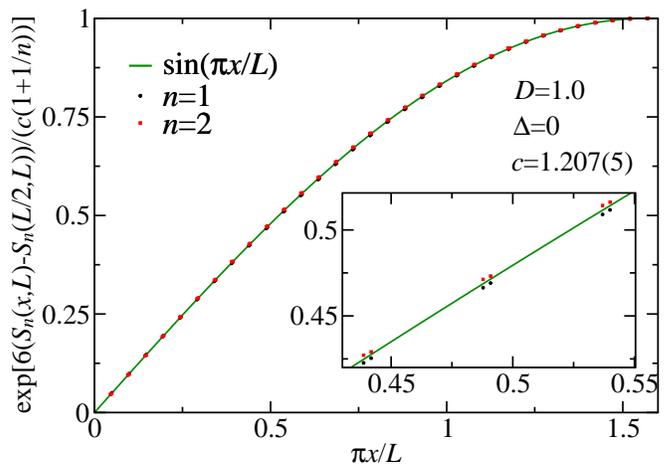} 
\par\end{centering}

\caption{(Color online) The scaling function $f_{L}^{D}\left(x\right)$ measured
by using the Rényi entropies $S_{1}$ and $S_{2}$ {[}see Eq.\ (\ref{eq:scaling-f}){]}
for chains of anisotropy $\Delta=0$ and lattice size $L=2^{10}$
averaged over $145\,000$ samples.\label{fig:Chord-length}}
\end{figure}

The chord length is another hallmark of finite-size scaling functions
of conformally invariant systems. It is intriguing that our numerical
results are consistent with a scaling function that is also the chord
length $f_{L}^{{\rm CFT}}\left(x\right)=\frac{L}{\pi}\sin\left(\frac{\pi}{L}x\right)$.
Since this may inspire new analytical insights in those correlated
disorder systems, we are now going to verify this statement with higher
precision in the case of the Rényi entanglement entropy. We are restricted
to the $\Delta=0$ case since we can deal with quite long chains,
via a mapping to a free fermion system. In these long chains, we neglect
the subleading terms in Eq.\ (\ref{eq:CFT-entanglement}). In Fig.\ \ref{fig:Chord-length},
we plot 
\begin{equation}
\frac{L}{\pi}f_{L}^{D}\left(x\right)=\exp\left(6\frac{S_{n}\left(x,L\right)-S_{n}\left(L/2,L\right)}{c_{D}^{{\rm eff}}\left(1+1/n\right)}\right),\label{eq:scaling-f}
\end{equation}
as a function of $\frac{\pi x}{L}$, for $n=1$ and $n=2$ and lattice
size $L=2^{10}$. The data are averaged over $145\,000$ disorder
realizations. We observe a clear agreement with the sine function
(continuous green line) $\sin\left(\frac{\pi x}{L}\right)$, implying
the leading chord length $f_{L}^{{\rm CFT}}$ dependence for the finite-size
scaling functions in our correlated random systems.

In the inset, we zoom in on the region where the curves most largely
disagree. Close inspection indicates that the tiny deviations from
$\sin\left(\pi x/L\right)$ are due to the imprecision (although small)
of the effective central charge value $c_{D}^{{\rm eff}}=1.207(5)$.
This may be attributed to the systematic error induced by the non-leading
terms in Eq.\ (\ref{eq:CFT-entanglement}). We also tried to fit
the data using other trial functions compatible with the symmetry
of $S_{n}$, namely, $f_{1}=\sin\left(\pi x/L\right)+A\sin^{3}\left(\pi x/L\right)$
and $f_{2}=\sin\left(\pi x/L\right)+B\sin\left(3\pi x/L\right)$.
The best fit gives $\left(A,B\right)=\left(-1.2,-1.1\right)\times10^{-3}$
and $\left(-1.5,-1.3\right)\times10^{-3}$ for $n=1$ and $2$, respectively,
which can be taken as zero within our numerical accuracy. Thus, the
chord length is likely the scaling function of $S_{n}$.

Moreover, as usually happens in conformally invariant systems, we
would like to verify if this same scaling function $f_{L}^{{\rm CFT}}$
also gives the leading behavior for other quantities such as the average
transversal spin-spin correlation function $C^{zz}(x,L)=\overline{\left\langle S_{i}^{z}S_{i+x}^{z}\right\rangle }$.

In order to show this, we recall that, to leading order, 
\[
C^{zz}(x,L)=-\left[Lg\left(x/L\right)\right]^{-\eta_{z}},
\]
for $x$ odd. For $x$ even, $C^{zz}=0$. Here, $g\left(z\right)$
is a positive periodic function of unity period and $g(z+1/2)=g(1/2-z)$;
and $\eta_{z}$ is the leading decay exponent. Indeed, for the clean
system, it is well known\ \cite{lieb-schultz-mattis} that 
\[
C^{zz}(x,L)=-\left(\pi f_{L}^{{\rm CFT}}\left(x\right)\right)^{-2},
\]
i.e., the decay exponent $\eta_{z}=2$ and the periodic function is
a simple sine. The same universal exponent $\eta_{z}=2$ happens in
the uncorrelated disorder case\ \cite{hoyosvieiralaflorenciemiranda}
$C^{zz}\sim-1/(12x^{2})$ for $1\ll x\ll L$. Unfortunately, the function
$g(z)$ is unknown in this case.

\begin{figure}
\begin{centering}
\includegraphics[clip,width=1\columnwidth]{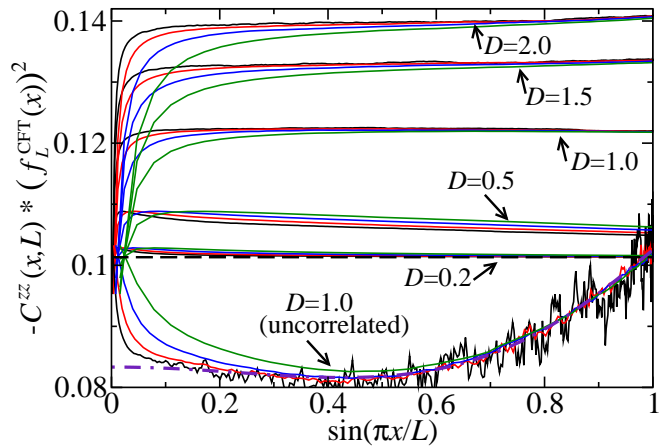} 
\par\end{centering}

\caption{(Color online) The transversal spin-spin correlation function for
various disorder strengths $D=0.2$, $0.5$, $1.0$, $1.5$, and $2.0$
(as indicated) and lattice sizes $L=200$ (green), $400$ (blue),
$800$ (red), and $1600$ (black). For comparison, we also plot $C^{zz}$
for the clean system (dashed line) and for the system with uncorrelated
disorder with $D=1.0$ (bottommost curves). Notice the stronger fluctuations
in this latter case. The dotted-dashed line is the fitting curve for
the uncorrelated case in the region $\sin(\pi x/L)>1/2$ (see text).
In all cases, the data are averaged over $10^{6}$ different disorder
configurations, except for $L=1\,600$ where the number of samples
used is $50\,000$. Error bars are not shown for clarity but are of
the order of the data fluctuations.\label{fig:Czz}}
\end{figure}

We plot in Fig.\ \ref{fig:Czz} $C^{zz}(x,L)\times\left(f_{L}^{{\rm CFT}}\left(x\right)\right)^{2}$
as a function of $\sin\frac{\pi x}{L}$, for several values of the
disorder strength $D$ and for lattice sizes $L=200,\,400,\,800$,
and $1600$. The fact that this combination saturates to a constant
for $x\gtrsim x_{{\rm min}}(L)$ indicates that the chord length,
for all the considered values of $D$, is the finite-size scaling
function for our correlated disorder case. For $x\lesssim x_{{\rm min}}(L)$,
finite-size corrections to scaling are expected {[}notice $x_{{\rm min}}(L\rightarrow\infty)\rightarrow0${]}.
This is in contrast with the uncorrelated disorder case, as shown
(for comparison) in Fig.\ \ref{fig:Czz}. It is clear in this case
that the chord length is not the finite-size scaling function. This
is not a surprise since it is already known that the finite-size scaling
function of the entanglement entropies $S_{n}$ is not a simple sine\ \cite{fagotti-calabrese-moore-prb11}.
Furthermore, we also verified that the scaling function appearing
in $S_{n}$ found in Ref.\ \onlinecite{fagotti-calabrese-moore-prb11}
is not the scaling function appearing in $C^{zz}$. Namely, a fitting
for the periodic function $g(z)$ in the asymptotic region $\sin(\pi z)>1/2$
and for $L=800$ is given by 
\[
g(z)=\frac{1}{\pi}\sum_{n=1}^{4}A_{2n-1}\sin((2n-1)\pi z),
\]
with $A_{1}=1.038$, $A_{3}=0.040$, $A_{5}=-0.007$, and $A_{7}=-0.003$,
with similar results for $L=1\,600$.

The fact that the chord length $f_{L}^{{\rm CFT}}$ is the scaling
function for both the entropies and the correlation function in our
correlated disorder case is an additional resemblance to the long-distance
behavior of conformally invariant systems. Actually, for $D<D_{c}$,
even the prefactor of the correlation seems to be the same as the
one of the clean system (see the dashed line in Fig.\ \ref{fig:Czz}
and compare with the case $D=0.2<D_{c}$).

Finally, we comment on the remarkable similarity between the Shannon
mutual information and the Rényi entanglement entropy $S_{2}$ as
shown in Fig.\ \hyperref[fig:Entanglement-entropy]{\ref{fig:Entanglement-entropy}(a)}.
As recently conjectured, the Shannon mutual information behaves like
$S_{2}$ in the long-length limit $1\ll x\ll L$ for clean conformally
invariant systems\ \cite{alcaraz-rajabpour-prl13,alcaraz-rajabpour-prb14,stephan-prb14}.
It is remarkable that the same similarity is present in our random
spin chain. Evidently, one would also like to compare with the mutual
information of spin chains governed by infinite-randomness fixed points.
Although $I$ is not computed in the literature, this is a simple
task which we accomplish in Appendix\ \ref{sec:App}. It turns out
that $I\left(x,L\right)=S_{n}\left(x,L\right)$ for random singlet
states (and for any $n$), which trivially agrees with the conjecture.

In summary, our numerical results indicate that the ground states
of quantum chains governed by this line of fixed points have entanglement
properties which are similar to those of conformally invariant ground
states. This is an unexpected result which certainly deserves further
attention in order to understand the complete generality of the results
derived from conformally invariant systems.

\section{Strong-disorder renormalization group\label{sec:SDRG}}

In this section we provide a strong-disorder renormalization-group
(SDRG) treatment of our correlated disorder spin chain in Eq.\ (\ref{eq:Hamiltonian}).
The main purpose is to understand the role of the anisotropy in our
line of finite-disorder fixed points. As shown below, this gives us
support to the conjecture that the anisotropy is an irrelevant parameter
in the region $-1<\Delta\leq1$. In addition, this treatment also
sheds some light on the nature of the ground state wave function of
our system.

The SDRG method\ \cite{MDH-PRL,MDH-PRB,bhatt-lee} gives us an appropriate
machinery for studying infinite-randomness fixed points. It gives
us an asymptotically exact description of the low-energy eigenlevels
of the system\ \cite{igloi-review}. In addition, it has also been
used to describe systems governed by finite-disorder fixed points,
giving us plausible results\ \cite{westerberg-PRB-FM-AF,yusuf-zigzag,hoyos-ladders}.
Up to now, a precise comparison of the SDRG results derived in this
latter case with exact results is still lacking. In the present section,
we are going to present such a comparison for our correlated random
system.

The main idea of the SDRG method is to integrate out local high-energy
degrees of freedom, renormalizing the remaining ones via perturbation
theory. In this way the low-energy physics is accessed, provided that
the perturbative procedure becomes more accurate along the renormalization-group
flow.

Here, our purpose is to study the perfectly correlated case of Eq.\ (\ref{eq:Hamiltonian}).
However, because new operators arise along the RG flow, it is more
convenient to expand the parameter space and consider the more general
Hamiltonian 
\begin{equation}
H=\sum_{i}H_{i}=\sum_{i}J_{i}\left(S_{i}^{x}S_{i+1}^{x}+S_{i}^{y}S_{i+1}^{y}+\Delta_{i}S_{i}^{z}S_{i+1}^{z}\right).\label{eq:H2}
\end{equation}
There are two differences from Eq.\ (\ref{eq:Hamiltonian}): (i)
the anisotropy $\Delta_{i}$ is now a random variable, and not fixed
as before, and is correlated with the coupling constants $J_{i}$
as we are going to explain below. (ii) The coupling constants $J_{i}$
are not \emph{perfectly} correlated (as $J_{1},J_{1},J_{3},J_{3},\dots$).

The local energy scale is $\epsilon_{i}=J_{i}(1+\Delta_{i})/2$ which
is the local gap of $H_{i}$. As the SDRG method is an energy-based
method, what is relevant is the correlation among the random scales
$\epsilon_{i}$. For this reason, we quantify the correlation among
the random scales of the system via the quantity 
\begin{equation}
\alpha=\alpha_{{\rm even,odd}}+\alpha_{{\rm odd,even}},\label{eq:correlation}
\end{equation}
where $\alpha_{{\rm even,odd}}=(\left\langle \epsilon_{2i}\epsilon_{2i+1}\right\rangle -\epsilon_{{\rm odd}}\epsilon_{{\rm even}})/\sigma_{{\rm odd}}\sigma_{{\rm even}}$,
$\epsilon_{{\rm x}}=\overline{\epsilon_{{\rm x}}}$ is the disorder
average over ${\rm x}={\rm even,odd}$ sites, and $\sigma_{{\rm x}}^{2}=\overline{\epsilon_{{\rm x}}^{2}}-\overline{\epsilon_{{\rm x}}}^{2}$
is the variance. Likewise $\alpha_{{\rm odd,even}}=(\left\langle \epsilon_{2i-1}\epsilon_{2i}\right\rangle -\epsilon_{{\rm odd}}\epsilon_{{\rm even}})/\sigma_{{\rm odd}}\sigma_{{\rm even}}$
measures the correlation between the odd-numbered sites and their
rightmost neighbors. Notice that for uncorrelated disorder $\alpha=0$,
whereas $\alpha=1$ for the case of \emph{perfectly} correlated disorder,
as in Eq.\ (\ref{eq:Hamiltonian}).

\begin{figure}
\begin{centering}
\includegraphics[clip,width=1\columnwidth]{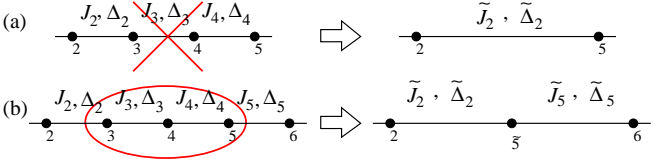} 
\par\end{centering}

\caption{(Color online) Decimation procedure for (a) uncorrelated and (b) correlated
disorder.\label{fig:Decimation}}
\end{figure}

It is instructive to review the SDRG procedure applied to the uncorrelated
case $\alpha=0$ in the region $-1<\Delta_{i}\leq1$\ \cite{fisher94-xxz}.
One searches for the greatest local energy scale $\Omega=\max\{\epsilon_{i}\}$,
say $\Omega=\epsilon_{3}$, and then treats $H_{2}+H_{4}$ as a perturbation
to $H_{3}$. As a result, spins $S_{3}$ and $S_{4}$ become locked
in a singlet and the neighboring spins $S_{2}$ and $S_{5}$ experience
an effective interaction given by $\tilde{H}_{2}=\tilde{J}_{2}\left(S_{2}^{x}S_{5}^{x}+S_{2}^{y}S_{5}^{y}+\tilde{\Delta}_{2}S_{2}^{z}S_{5}^{z}\right)$,
where the renormalized couplings are $\tilde{J}_{2}=\frac{J_{2}J_{4}}{J_{3}\left(1+\Delta_{3}\right)}$
and $\tilde{\Delta}_{2}=\frac{1}{2}\Delta_{2}\Delta_{4}\left(1+\Delta_{3}\right)$
{[}see Fig.\ \hyperref[fig:Decimation]{\ref{fig:Decimation}(a)}{]}.
Within this procedure\ \cite{fisher94-xxz}, it was shown that the
fixed point is of infinite-disorder type, i.e., $\sigma_{\epsilon}/\overline{\epsilon}\rightarrow\infty$
when $\Omega\rightarrow0$ (which provides \emph{a posteriori} justification
of the perturbative procedure), and that the anisotropy always renormalizes
to zero (except for the isotropic case $\Delta_{i}=1$). This conclusion
is valid whenever $-\frac{1}{2}<\Delta_{i}<1$. For $-1<\Delta_{i}<-\frac{1}{2}$,
the same conclusion applies only if the bare disorder of the system
is sufficiently strong. For $\left|\Delta_{i}\right|>1$, the long-range
order of the system is not changed by disorder, i.e., the system keeps
its ferromagnetic (for $\Delta_{i}<-1$) or antiferromagnetic (for
$\Delta_{i}>1$) Ising character. The entanglement properties are
thus similar to that of the clean system in which a finite correlation
length exists, i.e., it obeys the usual area law.

Clearly, this SDRG procedure cannot be applied for correlated disorder
$\alpha\neq0$ because the assumption that $\Omega=\epsilon_{3}\gg\epsilon_{2,4}$
fails. This can be circumvented by choosing a larger spin cluster
that also includes all the possible strong local couplings. Because
the correlations in Eq.\ (\ref{eq:Hamiltonian}) are short ranged,
such a cluster will be composed of three spins. For instance if $\Omega=\epsilon_{3}$,
then spins $S_{3}$ and $S_{4}$ belong to that triad of spins. In
order to decide if the third spin is either $S_{2}$ or $S_{5},$
we look for $\max\{\epsilon_{2},\epsilon_{4}\}$. In fact, a more
precise measure of the local energy scale $\varepsilon_{i}$ is the
mass gap of $H_{i}+H_{i+1}$. Thus, we redefine the SDRG cutoff energy
scale to $\Omega=\max\{\varepsilon_{i}\}$. This identifies the strongly
correlated spin triad. Once the energy cutoff is found, say $\Omega=\varepsilon_{3}$,
the next step is to treat $H_{2}+H_{5}$ as a perturbation to $H_{3}+H_{4}$.
Since the ground state of $H_{3}+H_{4}$ is a doublet it can be recast
as an effective spin-1/2 degree of freedom. Thus, projecting $H_{2}+H_{5}$
onto this doublet is equivalent to replacing spins $S_{3}$, $S_{4}$,
and $S_{5}$ by an effective spin $\tilde{S}_{5}$, which is connected
to $S_{2}$ and $S_{6}$ via effective couplings as depicted in Fig.\ \hyperref[fig:Decimation]{\ref{fig:Decimation}(b)}
(see details in Appendix\ \ref{sec:3-spins}).

Notice the fundamental difference between these two approaches. For
uncorrelated disorder, the ground state is always a random singlet
state regardless of the details of the random variables. For correlated
disorder, on the other hand, the effective spin $\tilde{S}_{5}$ has
strong correlations with the original spins $S_{3}$, $S_{4}$, and
$S_{5}$ if $\epsilon_{3}\approx\epsilon_{4}$. Therefore, this correlation
can be transmitted away to the rest of the chain when $\tilde{S}_{5}$
is decimated out in a later stage of the SDRG flow. In the end, correlations
among the original spins can be greatly enhanced and the ground state
can be fundamentally different from the random singlet state. This
is indeed the case as shown below.

Unfortunately, the effective couplings $\tilde{J}_{i}$ and $\tilde{\Delta}_{i}$
cannot be worked out analytically in an easy way (see Appendix\ \ref{sec:3-spins}).
We have implemented the SDRG decimation procedure numerically. However,
before presenting our numerical results, it is instructive to consider
some limiting cases of interest which can shed some light about the
renormalized system.

Let us first consider the case of perfectly correlated disorder $J_{4}=J_{3}$
and $\Delta_{4}=\Delta_{3}$ {[}which is exactly the one we are interested
in Eq.\ (\ref{eq:Hamiltonian}){]}. Projecting $H_{2}+H_{5}$ onto
the ground state of $H_{3}+H_{4}$ yields an effective Hamiltonian
as depicted in Fig.\ \hyperref[fig:Decimation]{\ref{fig:Decimation}(b)}
with 
\begin{equation}
\tilde{J}_{i}=\frac{2J_{i}}{\sqrt{8+\Delta_{3}^{2}}}\mbox{ and }\tilde{\Delta}_{i}=\left(\frac{\Delta_{3}+\sqrt{8+\Delta_{3}^{2}}}{4}\right)\Delta_{i}.\label{eq:eff-JD}
\end{equation}
Two aspects of the renormalized couplings are noteworthy. (i) Even
if the bare system has perfectly correlated disorder as in Eq.\ (\ref{eq:Hamiltonian})
(i.e., the couplings are $J_{1}J_{1}J_{3}J_{3}J_{5}J_{5}\dots$),
the renormalized system does not display such a feature: $J_{1}\tilde{J}_{1}\tilde{J}_{5}J_{5}\dots$.
This is the reason why we have expanded the parameter's space of our
original Hamiltonian. (ii) The renormalized anisotropies are smaller
than the original ones. This suggests that, as for the uncorrelated
disorder case, the anisotropy renormalizes to zero. Naturally, both
points (i) and (ii) have been extensively tested as shown below.

Another case that can be studied exactly is the free-fermion one $\Delta_{i}=0$
(see Appendix\ \ref{sec:App}). Here, the $\Delta_{i}$ remains zero
along the SDRG flow and the renormalized coupling constants in Fig.\ \hyperref[fig:Decimation]{\ref{fig:Decimation}(b)}
are 
\begin{equation}
\tilde{J}_{2}=\frac{J_{2}J_{4}}{\sqrt{J_{3}^{2}+J_{4}^{2}}}\mbox{ and }\tilde{J}_{5}=\frac{J_{3}J_{5}}{\sqrt{J_{3}^{2}+J_{4}^{2}}}.\label{eq:eff-JJ-freefermion}
\end{equation}
Again, the perfect correlations of the original chain $J_{1}J_{1}J_{3}J_{3}J_{5}J_{5}\dots$
are lost after just the first decimation.

Finally, the third case in which the decimation rules can be worked
out analytically is the isotropic $SU(2)$ one where $\Delta_{i}=1$.
As in the free-fermion case, the anisotropy does not change and is
fixed at $\tilde{\Delta}_{i}=1$. The effective coupling constants
are 
\begin{equation}
\tilde{J}_{2}=\left(\frac{2J_{4}-J_{3}+\gamma}{3\gamma}\right)J_{2}\mbox{ and }\tilde{J}_{5}=\left(\frac{2J_{3}-J_{4}+\gamma}{3\gamma}\right)J_{5},\label{eq:eff-JJ-isotropic}
\end{equation}
with $\gamma=\sqrt{J_{3}^{2}-J_{3}J_{4}+J_{4}^{2}}$. Again, as in
the previous cases, the perfect correlation is lost along the SDRG
flow.

The lessons learned from this little digression (and confirmed by
our numerical study below) are (i) the fixed points are either $\tilde{\Delta}_{i}=0$
or $\tilde{\Delta}_{i}=1$ and (ii) perfect correlation on the energy
scales is lost during the SDRG flow. Lesson (ii) raises an important
question. If perfect correlation is lost, does the system flow to
a totally uncorrelated fixed point which would be of infinite-randomness
type? As shown below, this is indeed the case \emph{except} for the
free-fermion case $\Delta_{i}=0$.

We now report our numerical study of the SDRG method on our random
spin chains with correlated disorder Eq.\ (\ref{eq:Hamiltonian}).
The SDRG decimation procedure (as described above) is depicted in
Fig.\ \hyperref[fig:Decimation]{\ref{fig:Decimation}(b)}. As the
energy scale $\Omega$ is lowered, we keep track of the renormalized
values of the disorder correlation measure $\tilde{\alpha}$ and the
mean $\left\langle \tilde{\Delta}\right\rangle $ anisotropy, together
with its variance $\sigma_{\tilde{\Delta}}^{2}$, and the length scale
$\xi=\rho^{-1}$ which is also the inverse of density $\rho=L/n_{\Omega}$
($n_{\Omega}$ being the number of spin clusters at the energy scale
$\Omega$). In all of our runs, the chain is decimated until the system
is reduced to 20 spin clusters.

\begin{figure}[t]
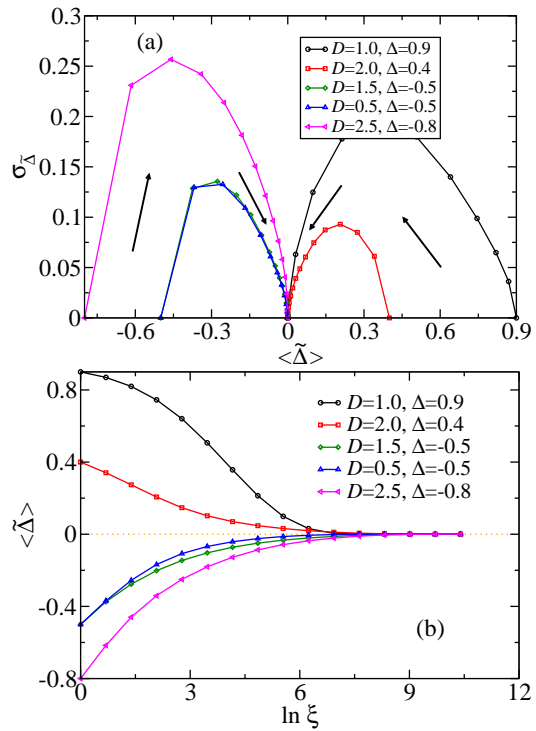

\begin{centering}
\includegraphics[clip,width=0.8\columnwidth]{fig9a}\\
 \includegraphics[clip,width=0.8\columnwidth]{fig9b} 
\par\end{centering}

\caption{(Color online) The anisotropy along the SDRG flow for various disorder
parameters $D$ and initial conditions $\Delta$, for the case of
perfectly correlated disorder $\alpha=1$. In panel (a) the standard
deviation $\sigma_{\tilde{\Delta}}$ is plotted as a function of the
average $\left\langle \tilde{\Delta}\right\rangle $, with the arrows
indicating the direction of the SDRG flow. In panel (b) $\left\langle \tilde{\Delta}\right\rangle $
is plotted as a function of the SDRG length scale $\xi$. The lines
are guides to the eyes and the data are averaged over only $10$ disorder
realizations of a large lattice size of $L=10^{6}$.\label{fig:Anisotropy}}
\end{figure}

We show in Fig.\ \ref{fig:Anisotropy} that the anisotropy is an
irrelevant perturbation at the critical line $-1<\Delta<1$, i.e.,
the low-energy physics is governed by the free-fermion fixed point
$\tilde{\Delta}=0$. In this figure, only a few disorder realizations
are enough for a reasonable precision due to the very large lattice
size used.

\begin{figure}
\begin{centering}
\includegraphics[clip,width=0.8\columnwidth]{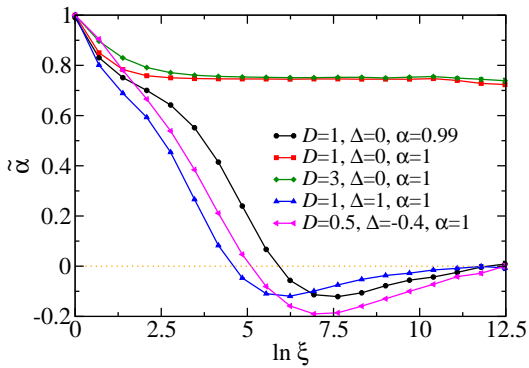} 
\par\end{centering}

\caption{(Color online) Disorder correlation along the SDRG flow for various
disorder strengths $D$ and anisotropies $\Delta$. The correlation
parameter $\alpha$ is plotted as a function of the SDRG length scale
$\xi$. Data are averaged over $1\,000$ disorder realizations of
large chains of size $10^{7}$. Lines are guides to the eyes.\label{fig:correlation}}
\end{figure}

The effective disorder correlations of our system are shown in Fig.\ \ref{fig:correlation}.
We plot the correlation measure of the renormalized disorder variables
$\tilde{\alpha}$ along the SDRG flow, as a function of the length
scale $\xi$. As can be seen, except for the case where $\Delta=0$
\emph{and} \emph{ }perfectly correlated disorder $\alpha=1$, the
disorder correlation of the renormalized coupling constants $\tilde{\alpha}\rightarrow0$
at the final stages of the flow. This means that the true fixed point
is the one of the uncorrelated disorder case which is of infinite-randomness
type. Notice, however, that the correlation does not vanish fast.
In the RG language, this means a long incursion of the flow near the
fixed point of the system with perfectly correlated disorder. As a
consequence, the physics of relatively small system sizes is governed
by the finite-disorder fixed point. This is compatible with the results
we presented in Secs.\ \ref{sec:Dynamical-exponent} and \ref{sec:Entanglement}
for the $\Delta\neq0$ cases.

How can we explain the different behaviors between the cases $\Delta\neq0$
and $\Delta=0$? Actually, we have solid arguments in favor of a vanishing
random mass term only in the special case $\alpha=1$ and $\Delta=0$\ \cite{hoyos-etal-epl11}.
The results of Secs.\ \ref{sec:Dynamical-exponent} and \ref{sec:Entanglement},
derived for small lattice sizes $L\leq20$, are compatible with a
similar vanishing random mass for $\Delta\neq0$. However, as the
SDRG results reveal (see Fig.\ \ref{fig:correlation}), there is
a relatively large crossover length above which random mass becomes
relevant.

\begin{figure}[b]
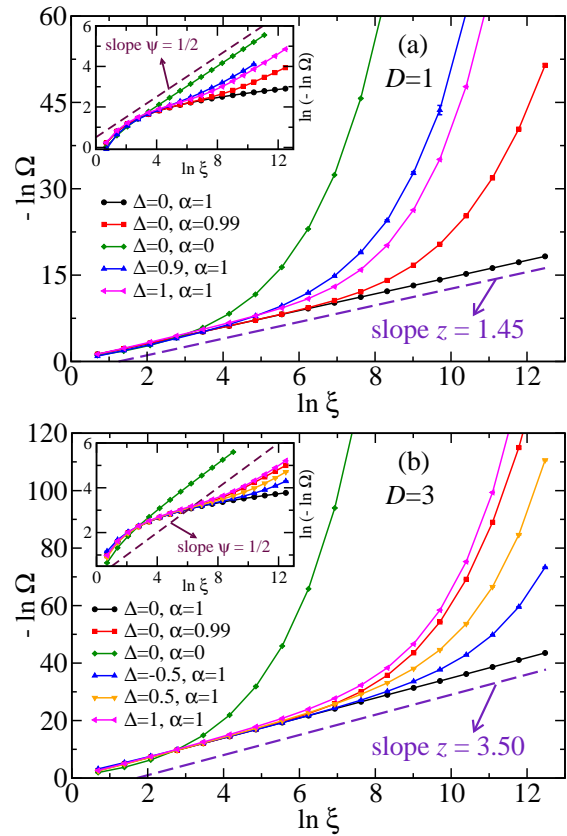

\begin{centering}
\includegraphics[clip,width=0.85\columnwidth]{fig11a}\\
 \includegraphics[clip,width=0.85\columnwidth]{fig11b} 
\par\end{centering}

\caption{(Color online) The relation between the energy $\Omega$ and the length
$\xi$ scales for the disorder strengths (a) $D=1$ and (b) $D=3$
and various anisotropies $\Delta$ and correlation $\alpha$. Dashed
lines are the best fits for the case $\Delta=0$ and $\alpha=1$ restricted
to $\ln\xi\geq7$. Inset: The same as in the main plot with the logarithmic
value of the vertical axis. Dashed lines are the corresponding theoretical
slopes for infinite-randomness physics.\label{fig:scaling}}
\end{figure}

In order to test this interpretation, we investigate the system dispersion
relation by plotting the energy scale $\Omega$ as a function of the
length scale $\xi$. For conventional dynamics, a power-law scaling
$\Omega\sim\xi^{-z}$ is expected, compatible with a finite-disorder
fixed point (absence of random mass). On the other hand, for an infinite-randomness
fixed point (presence of random mass)\ \cite{fisher94-xxz}, an activated
dynamical scaling takes place in which $\ln\Omega\sim-\xi^{\psi}$,
with universal tunneling exponent $\psi=\frac{1}{2}$. Therefore,
for $\Delta\ne0$ (irrespective whether disorder is perfectly correlated
or not), or for $\Delta=0$ but $\alpha\neq1$ (disorder not perfectly
correlated), we expect a conventional power-law scaling in the earlier
stages of the SDRG flow, and activated dynamics after a crossover.
This is indeed the observed scenario as shown in Fig.\ \ref{fig:scaling}.
Notice that, for small length scales, a power-law scaling $\Omega\sim\xi^{-z}$
takes place. This regime is governed by the ``unstable'' fixed point
which has no random mass. In this transient regime, the critical dynamical
exponent does not depend on the anisotropy $\Delta$, in agreement
with the results obtained by exact diagonalization in Sec.\ \ref{sec:Dynamical-exponent}.

Interestingly, the SDRG method here presented can be quantitatively
compared with exact results. The dynamical exponent $z$, obtained
from the fits in Fig.\ \ref{fig:scaling}, is plotted in Fig.\ \ref{fig:Finite-size-effects}
together with the expected exact value. Two features are noteworthy.
(i) For weak disorder $D<D_{c}$ the SDRG method indicates a universal
finite-disorder fixed point. Because the corresponding $z_{{\rm SDRG}}$
is less than $z_{{\rm clean}}=1$, we interpret that the correct fixed
point is the one of the clean system. This is a consequence of the
delocalized modes (spinons) are energetically favorable against the
localized ones obtained by the SDRG method; i.e., the delocalized
modes are the true low-energy modes. (ii) The dynamical critical exponent
$z_{{\rm SDRG}}$ is quantitatively close to the exact one even for
small $z,$ say $z\approx1.5$. Thus, even though the SDRG is only
justifiable at strong disorder ($z\gg1$), it already gives relatively
accurate results even for moderate $z$. In addition, notice that
the $z_{{\rm SDRG}}$ is always smaller than $z$. We interpret this
in the following manner. Let us suppose we could untangle the localized
and delocalized modes coming from the limits of weak and strong disorder,
respectively. Clearly, in the way the SDRG method is formulated, only
the localized modes are captured, namely, resonances of a triad of
strongly coupled spin clusters. However, these are not the true modes
of the system. There are certain aspects of the delocalized theory
that are not taken into account. For instance, our spin-triad resonance
may interact with spinons. This interaction would further lower the
energy of the modes, which would correspond to a larger dynamical
exponent. Hence $z$ is greater than or equal to $z_{{\rm clean}}=1$
and $z_{{\rm SDRG}}$.

\begin{figure}
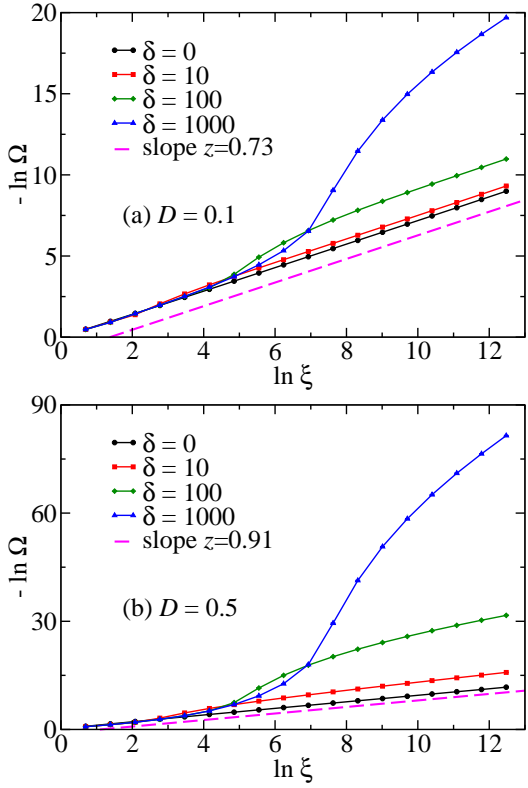

\begin{centering}
\includegraphics[clip,width=0.8\columnwidth]{fig12a}\\
 \includegraphics[clip,width=0.8\columnwidth]{fig12b} 
\par\end{centering}

\caption{(Color online) The scaling relation between energy $\Omega$ and length
$\xi$ for the $\Delta=0$ chain with the even coupling constants
spatially shifted by an amount $\delta$ and disorder strengths (a)
$D=0.1$ and (b) $D=0.5$. Dashed lines are the fits for the unshifted
case for $\ln\xi\geq7$. \label{fig:scaling-shifted}}
\end{figure}

Let us now briefly discuss on the possibility that the observed finite
random mass (in the case $\Delta\neq0$) is just an artificial product
of the inexactness of the SDRG method. One could argue that our decimation
procedure does not preserve locally the disorder correlation, and
therefore random mass is introduced even for $\Delta=0$. This would
imply that the nearly perfect match between $z$ and $z_{{\rm SDRG}}$
is just a coincidence. The important detail that must be kept in mind
is that the random mass is a coarse-grained concept, and hence the
local constraint $\tilde{J}_{2i-1}=\tilde{J}_{2i}$ is not necessary.

\begin{figure}
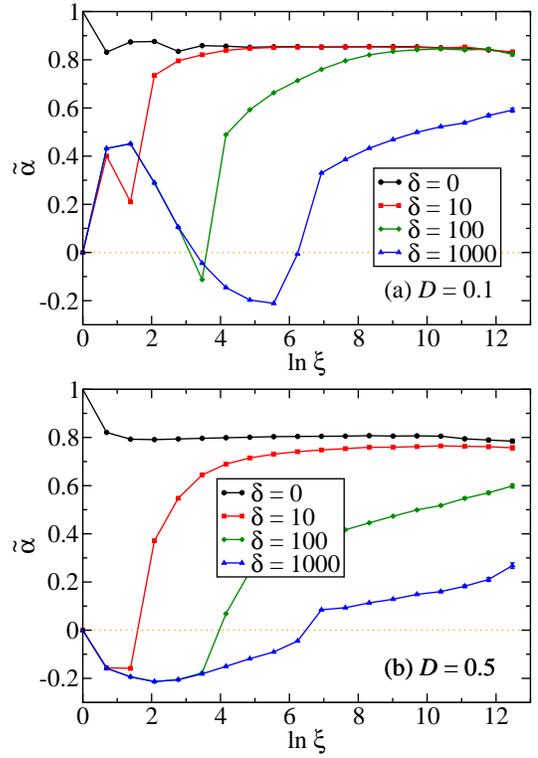

\begin{centering}
\includegraphics[clip,width=0.8\columnwidth]{fig13a}\\
 \includegraphics[clip,width=0.8\columnwidth]{fig13b} 
\par\end{centering}

\caption{(Color online) The correlation measure $\tilde{\alpha}$ for the $\Delta=0$
chain with the even coupling constants shifted by an amount $\delta$
and disorder strengths (a) $D=0.1$ and (b) $D=0.5$.\label{fig:correlation-shifted}}
\end{figure}

We believe that indeed the absence of the random mass is captured
by the SDRG method even when it is not present in the short length
scale. This reasoning is based on the following study. Instead of
using the correlated chain in Eq.\ (\ref{eq:Hamiltonian}) where
the coupling constants at sites $2i-1$ and $2i$ are correlated ($\dots J_{1}J_{1}J_{3}J_{3}J_{5}J_{5}\dots$),
we shift the correlated couplings by an integer $\delta>2$ such that
the sequence of couplings is now $\dots J_{1}J_{1+\delta}J_{3}J_{3+\delta}J_{5}J_{5+\delta}\dots$.
In this way, since $J_{i}$ has no correlation with $J_{i+\delta}$
for $\delta>2$, there is no cancellation of the random mass term
at sites $i$ and $i+1$. The absence of random mass would be observed
only at length scales $\xi\gg\delta$, where the condition $\prod J_{2i}=\prod J_{2i-1}$
becomes approximately fulfilled. This is indeed what is verified in
the energy-length dispersion relation as shown in Fig.\ \ref{fig:scaling-shifted}.
For $\xi<\delta$, the system ``sees'' local random masses and the
scaling is activated. Only for $\xi\gg\delta,$ does the true asymptotic
regime take place where the dynamical critical exponent is finite
and equals the one of the unshifted ($\delta=0$) chain. Even the
correlation between the effective coupling constants becomes finite,
as plotted in Fig.\ \ref{fig:correlation-shifted}. For $\delta>2$,
the correlation $\alpha$ is initially zero, and after coarse-graining
it builds up as in the $\delta=0$ case. For large $\delta$, the
crossover from the activated to the power-law dynamical scaling is
very slow, taking over on a large range of length scales $\xi.$

Thus, it is plausible to conclude that our SDRG decimation procedure
does not introduce random masses in the $\Delta\neq0$ case, as happens
for the $\Delta=0$ case.

Finally, in order to conclude this section, let us discuss on a possible
generalization of the $c$ theorem \cite{zamolodchikov-c-theorem},
which states that if RG flow between two conformally invariant fixed
points $a$ and $b$ is in the direction $a\rightarrow b$, then the
corresponding central charges are such that $c_{a}\geq c_{b}$. In
the earlier years of the study of entanglement properties of random
spin chains, the quest for an effective $c$ theorem fostered intense
research \cite{refael-moore-07,refael-moore-prl04,santachiara-06,fidkowski-etal-prb08}.
It turned out that such generalization was proven impossible for spin
chains governed by infinite-randomness fixed points since violations
were observed \cite{santachiara-06,fidkowski-etal-prb08}. It is worth
asking if such violation also happens in our correlated disordered
case.

Probing the SDRG flow (as described in Sec.\ \ref{sec:SDRG}) for
the XX model ($\Delta_{i}=0$) described by the Hamiltonian (\ref{eq:H2}),
we conclude that there is no violation of the $c$ theorem. This conclusion
stems from the following reasoning. Consider the SDRG flow in the
$D$-$\alpha$ parameter space, where the disorder strength $D$ is
defined in Eq.\ (\ref{eq:P(J)}) and the correlation in the disorder
variables $\alpha$ is defined in Eq.\ (\ref{eq:correlation}). As
shown in Fig.\ \ref{fig:correlation}, any perturbation away from
the perfectly correlated disorder drives the flow towards the uncorrelated
disorder case. Thus, the line of finite-disorder fixed points is unstable
towards the infinite-disorder fixed point along the $\alpha$ direction.
As shown in Ref.\ \onlinecite{hoyos-etal-epl11} (see also Fig.\ \ref{fig:dynamical-exponent}),
weak disorder is perturbatively relevant in the perfectly correlated
case ($\alpha=1$), hence the RG flow goes from the finite-disorder
fixed points towards the clean fixed point; i.e., the line of fixed
points is unstable along the $D$ direction as well. Finally, it is
well know that the clean fixed point is unstable towards the infinite-disorder
fixed point along the $D$ direction (as long as $\alpha\neq1$).
As $c_{{\rm IFRP}}^{{\rm eff}}=\ln2<c_{{\rm clean}}=1<c_{D}^{{\rm eff}}$,
we conclude that the SDRG flow is compatible with an ``effective''
$c$ theorem. At the line of finite-disorder fixed points, one may
inquire if there is any violation of this effective theorem. Notice
that there is no flow among these fixed points since their basin of
attractions is null.

\section{Conclusion and discussions\label{sec:Conclusion}}

We have studied the critical spin-1/2 XXZ chain with short-range correlated
disorder. Due to absence of the random mass, this system has the interesting
property of realizing a line of finite-disorder fixed points tuned
by the disorder strength. Such mass absence is asserted by the local
correlation $J_{2i}=J_{2i-1}$ in the free-fermion case $\Delta=0$.
In the general case $\Delta\neq0$, the random mass is present although
its magnitude is quite small. Thus, the line of fixed points governs
the physics in the intermediate-energy regime. Here, we did not consider
the question of whether it is possible to devise local correlations
among the disorder variables of the microscopic model in Eq.\ (\ref{eq:Hamiltonian})
for $\Delta\neq0$ in order to ensure an effective theory free of
random mass.

As it is well known, the violation of the area law of the entanglement
entropy of critical systems happens for both clean and disordered
systems. The most studied disordered chains are those with uncorrelated
disorder which realizes infinite-randomness fixed points. In this
case, differently from the clean systems which are usually conformally
invariant, the $n$-Rényi entanglement entropies are independent of
the index $n$, being equal to the Shannon entanglement entropy.

In this paper, we performed an extensive study of the entanglement
properties of ground states of quantum chains governed by finite-disorder
fixed points. Surprisingly, we obtained quite different results from
those of infinite-disorder fixed points. In contrast, the leading
finite-size behaviors show striking resemblances with those of the
clean conformally invariant systems, even though the critical random
system under consideration is \emph{not} conformally invariant. These
resemblances are (a) the \emph{same $n$ dependence} occurs in the
Rényi entanglement entropy $S_{n}$, (b) the \emph{same periodic}
\emph{scaling function} $f_{L}^{{\rm CFT}}(x)=\frac{L}{\pi}\sin\left(\frac{\pi x}{L}\right)$
is applicable for the entanglement properties as well as for the spin-spin
correlation functions, and (c) the Shannon mutual information $I(x)$
and the Rényi entanglement entropy $S_{2}(x)$ share the \emph{same
leading }behavior as conjectured for conformally invariant systems.

From the results in the literature up to now, the logarithmic behavior
of the entanglement entropies seems to be a general feature of one-dimensional
critical systems regardless of being disordered or not. However, the
periodic finite-size scaling function $f_{L}^{{\rm CFT}}(x)$ appearing
in the argument of this logarithmic dependence (also in the correlation
functions) is normally expected to be a signature of the underlying
conformal field theory governing the long-distance physics of the
system. Our results give us two possibilities: either this behavior
is a consequence of more general assumptions than conformal invariance,
or there is some emerging effective conformal symmetry in those correlated
disordered systems which is not obvious. This latter possibility is
quite intriguing since the dynamical critical exponent $z>1$. Emerging
symmetries in critical random systems have been recently reported\ \cite{quito-hoyos-miranda-prl14},
but are still poorly understood.

This paper also gives us the first extension of the conjecture stating
that $S_{2}(x)$ and $I(x)$, in conformally invariant systems and
computed on certain special basis, share the same leading asymptotic
behavior for large systems and subsystem sizes. It would be quite
interesting to test this conjecture in other disordered systems.
\begin{acknowledgments}
This work was supported by CAPES, FAPESP, and CNPq (Brazilian agencies). 
\end{acknowledgments}
\appendix

\section{Entanglement in random singlet phases\label{sec:App}}

For completeness, we derive the Shannon mutual information and the
Rényi entanglement entropy for the random singlet state of the antiferromagnetic
spin-1/2 chain.

It has been shown\ \cite{hoyos-rigolin,hoyosvieiralaflorenciemiranda,tran-bonesteel-prb11,pouranvari-yang-prb13}
that the random singlet state is well approximated by a collection
of singlet pairs 
\begin{equation}
\left|\psi\right\rangle =\bigotimes_{i=1}^{L/2}\left|s_{i}\right\rangle =\bigotimes_{a=1}^{N_{a}}\left|s_{a}\right\rangle \bigotimes_{b=1}^{N_{b}}\left|s_{b}\right\rangle \bigotimes_{ab=1}^{N_{ab}}\left|s_{ab}\right\rangle ,\label{eq:RS-state}
\end{equation}
where $\left|s_{i}\right\rangle =\frac{1}{\sqrt{2}}\left(\left|+-\right\rangle -\left|-+\right\rangle \right)$
is the singlet state of the $i$-th singlet pair. We divide these
singlets into three categories: those in which (\emph{a}) both spins
belong to region ${\cal A}$, (\emph{b}) both spins belong to region
${\cal B}$, and (\emph{ab}) one of the spins belongs to region ${\cal A}$
while the other one belongs to ${\cal B}$. They are denoted respectively
by $\left|s_{a}\right\rangle $, $\left|s_{b}\right\rangle $ and
$\left|s_{ab}\right\rangle $, and there are $N_{a}$, $N_{b}$ and
$N_{ab}$ of each. The constraint is that $N_{a}+N_{b}+N_{ab}=L/2$.

The reduced density matrix $\rho_{A}$ can be easily computed as 
\begin{equation}
\rho_{A}=\bigotimes_{a}\left|s_{a}\right\rangle \left\langle s_{a}\right|\bigotimes_{ab}\frac{1}{2}\left(\left|+\right\rangle \left\langle +\right|+\left|-\right\rangle \left\langle -\right|\right).
\end{equation}
Thus, the nonvanishing eigenvalues of $\rho_{A}$ come from the \emph{ab}
singlets. There are $2^{N_{ab}}$ degenerate eigenvalues equal to
$\lambda_{i}=2^{-N_{ab}}$, and therefore, the $n$-Rényi entanglement
entropy is 
\begin{equation}
S_{n}=\frac{\ln\left(\sum_{i}\lambda_{i}^{n}\right)}{1-n}=\frac{\ln\left(2^{N_{ab}-nN_{ab}}\right)}{1-n}=N_{ab}\ln2,
\end{equation}
which does not depend on the index $n$.

The Shannon mutual information can be computed in a similar way. Let
us start with the Shannon entropy of the entire system. The random-singlet
state in Eq.\ (\ref{eq:RS-state}) has $2^{L/2}$ different configurations,
all of them occurring with the same probability $2^{-L/2}$. Therefore
${\rm Sh}\left({\cal A}\cup{\cal B}\right)=-2^{L/2}\times2^{-L/2}\ln2^{-L/2}=\frac{L}{2}\ln2$.
Now, let us compute the Shannon entropy for subsystem ${\cal A}$.
It is easy to see that all possible configurations will appear with
the same probability. Thus, our task is to compute only the number
of different configurations. Due to the $N_{a}$ singlets inside region
${\cal A}$, there will be a contribution of $2^{N_{a}}$ configurations.
Moreover, we have to take into account the $N_{ab}$ singlets which
are shared by both subsystems. Each such pair has one spin in subsystem
${\cal A}$ which will appear with equal probability in the $\left|+\right\rangle $
and $\left|-\right\rangle $ states. Thus, the total number of configurations
is $2^{N_{a}+N_{ab}}$. Likewise for the subsystem ${\cal B}$, the
total number of configurations is $2^{N_{b}+N_{ab}}$. We are now
in position to compute the Shannon mutual information: 
\begin{equation}
I={\rm Sh}\left({\cal A}\right)+{\rm Sh}\left({\cal B}\right)-{\rm Sh}\left({\cal A}\cup{\cal B}\right)=N_{ab}\ln2,
\end{equation}
where we have used $N_{a}+N_{b}+N_{ab}=L/2$. Therefore, the leading
term of the Shannon mutual information equals the leading term of
the Rényi entanglement entropy for all $n$ for random-singlet states.
Notice that this result does not depend exactly on how the singlets
are distributed on the chain. This is important only to relate $N_{ab}$
with $x$. Moreover, this result can be generalized to any spin $S$.
One needs only to replace $\ln2$ by $\ln2S+1$.

Finally, we mention that the number of spin singlets belonging to
both subsystems ${\cal A}$ and ${\cal B}$ can be computed with the
methods in the literature\ \cite{refael-moore-prl04,hoyosvieiralaflorenciemiranda}.
Let $x$ be the ${\cal A}$ subsystem size. It is found that for $1\ll x\ll L$,
$N_{ab}=\frac{1}{3}\ln x$. With this, one recovers the known result
that $S_{n}=\frac{1}{3}c_{{\rm RS}}\ln x$, with the universal prefactor
$c_{{\rm RS}}=\ln2$.

\section{Three-spins SDRG\label{sec:3-spins}}

We wish to treat $H^{\prime}$ as a perturbation to $H$ where 
\begin{align*}
H= & J_{3}\left(\frac{S_{3}^{+}S_{4}^{-}+{\rm h.c.}}{2}+\Delta_{3}S_{3}^{z}S_{4}^{z}\right)\\
 & +J_{4}\left(\frac{S_{4}^{+}S_{5}^{-}+{\rm h.c.}}{2}+\Delta_{4}S_{4}^{z}S_{5}^{z}\right),
\end{align*}
and 
\begin{align*}
H^{\prime}= & J_{2}\left(\frac{S_{2}^{+}S_{3}^{-}+{\rm h.c.}}{2}+\Delta_{2}S_{2}^{z}S_{3}^{z}\right)\\
 & +J_{5}\left(\frac{S_{5}^{+}S_{6}^{-}+{\rm h.c.}}{2}+\Delta_{5}S_{5}^{z}S_{6}^{z}\right).
\end{align*}
We then need to diagonalize $H$. Due to conservation of the total
$z$ magnetization $S^{z}$, the eigenvalue problem for this Hamiltonian
can be reduced by solving a single $3\times3$ matrix. Two eigenvectors
follow straightforwardly (corresponding to $S^{z}=\pm\frac{3}{2}$):
$\left|+++\right\rangle $ and $\left|---\right\rangle $, being degenerate
with eigenenergy $E_{4}=\frac{1}{4}\left(J_{3}\Delta_{3}+J_{4}\Delta_{4}\right)$.
There are three eigenvectors obtained from the $S^{z}=+\frac{1}{2}$
sector whose corresponding matrix is 
\[
\mathbb{M}=\frac{1}{2}\left(\begin{array}{ccc}
\frac{+J_{3}\Delta_{3}-J_{4}\Delta_{4}}{2} & J_{4} & 0\\
J_{4} & \frac{-J_{3}\Delta_{3}-J_{4}\Delta_{4}}{2} & J_{3}\\
0 & J_{3} & \frac{-J_{3}\Delta_{3}+J_{4}\Delta_{4}}{2}
\end{array}\right),
\]
where the order $\left|++-\right\rangle $, $\left|+-+\right\rangle $,
and $\left|-++\right\rangle $ was used. The same result is obtained
in the $S^{z}=-\frac{1}{2}$ sector.

Since the system is antiferromagnetic ($-1<\Delta_{2,3}\leq1$), the
ground states are doubly degenerate and belong to the sectors $S^{z}=\pm\frac{1}{2}$.
The eigenvalues of $\mathbb{M}$ are the roots of a cubic polynomial,
which we denote by $E_{1,2,3}$ with $E_{1}<E_{2,3,4}$. The degenerate
ground states are $\left|\tilde{+}\right\rangle =a_{1}\left|++-\right\rangle +a_{2}\left|+-+\right\rangle +a_{3}\left|-++\right\rangle $
and $\left|\tilde{-}\right\rangle =-a_{1}\left|--+\right\rangle -a_{2}\left|-+-\right\rangle -a_{3}\left|+--\right\rangle $,
with $a_{i}\in\Re$. Once the doublet is obtained, our task is to
project $H^{\prime}$ onto the doublet $\left|\tilde{\pm}\right\rangle $;
i.e., we need to project $S_{3}$ and $S_{5}$ onto the doublet $\left|\tilde{\pm}\right\rangle $:
\[
S_{i}^{\alpha}\rightarrow\left(\begin{array}{cc}
\left\langle \tilde{+}\left|S_{i}^{\alpha}\right|\tilde{+}\right\rangle  & \left\langle \tilde{+}\left|S_{i}^{\alpha}\right|\tilde{-}\right\rangle \\
\left\langle \tilde{-}\left|S_{i}^{\alpha}\right|\tilde{+}\right\rangle  & \left\langle \tilde{-}\left|S_{i}^{\alpha}\right|\tilde{-}\right\rangle 
\end{array}\right).
\]
We then find that $S_{3}^{+}\rightarrow-2a_{1}a_{2}\tilde{S}_{5}^{+}$,
$S_{5}^{+}\rightarrow-2a_{3}a_{2}\tilde{S}_{5}^{+}$, $S_{3}^{z}\rightarrow\frac{1}{2}\left(a_{1}^{2}+a_{2}^{2}-a_{3}^{3}\right)\tilde{S}_{5}^{z}$,
$S_{5}^{z}\rightarrow\frac{1}{2}\left(-a_{1}^{2}+a_{2}^{2}+a_{3}^{3}\right)\tilde{S}_{5}^{z}$,
where $\tilde{S}_{5}^{+}=\left(\begin{array}{cc}
0 & 1\\
0 & 0
\end{array}\right)$ and $\tilde{S}_{5}^{z}=\frac{1}{2}\left(\begin{array}{cc}
1 & 0\\
0 & -1
\end{array}\right)$. Thus, we arrive at the effective Hamiltonian 
\begin{align}
\tilde{H}^{\prime}= & \tilde{J}_{2}\left(\frac{S_{2}^{+}\tilde{S}_{5}^{-}+{\rm h.c.}}{2}+\tilde{\Delta}_{2}S_{2}^{z}\tilde{S}_{5}^{z}\right)\nonumber \\
 & +\tilde{J}_{5}\left(\frac{\tilde{S}_{5}^{+}S_{6}^{+}+{\rm h.c.}}{2}+\tilde{\Delta}_{5}\tilde{S}_{5}^{z}S_{6}^{z}\right),\label{eq:effectiveH}
\end{align}
with the renormalized couplings 
\[
\tilde{J}_{2}=-2a_{1}a_{2}J_{2},\quad\tilde{J}_{5}=-2a_{2}a_{3}J_{5},
\]
\[
\tilde{\Delta}_{2}=\frac{-a_{1}^{2}-a_{2}^{2}+a_{3}^{2}}{2a_{1}a_{2}}\Delta_{2},\quad\tilde{\Delta}_{5}=\frac{a_{1}^{2}-a_{2}^{2}-a_{3}^{2}}{2a_{2}a_{3}}\Delta_{5}.
\]
This decimation procedure is depicted in Fig.\ \hyperref[fig:Decimation]{\ref{fig:Decimation}(b)}.
Notice that a global signal can be gauged out for all couplings since
we can choose a harmless global factor in $\left|\tilde{-}\right\rangle $.

Unfortunately, we could not solve the coefficients $a_{i}$ analytically
for the ground state in the generic case. Nevertheless, three important
limiting cases can be worked out analytically: (i) perfect correlation
$J_{4}=J_{3}$ and $\Delta_{4}=\Delta_{3}$, (ii) free fermions $\Delta_{i}=0$,
and (ii) isotropic Heisenberg $\Delta_{i}=1$.

For the special case where $J_{4}=J_{3}$ and $\Delta_{4}=\Delta_{3}$,
the eigenvalues of $\mathbb{M}$ are $E_{1,2}=-\frac{1}{4}J_{3}\left(\Delta_{3}\pm\sqrt{8+\Delta_{3}^{2}}\right)$
and $E_{3}=0$, and the ground states are $\left|\tilde{+}\right\rangle =J_{3}\left|++-\right\rangle +2E_{1}\left|+-+\right\rangle +J_{3}\left|-++\right\rangle $
and $\left|\tilde{-}\right\rangle =-J_{3}\left|--+\right\rangle -2E_{1}\left|-+-\right\rangle -J_{3}\left|+--\right\rangle $
which can be recast as an effective spin-1/2 degree of freedom (notice
$\left\langle \tilde{+}\left|S^{z}\right|\tilde{+}\right\rangle =-\left\langle \tilde{-}\left|S^{z}\right|\tilde{-}\right\rangle =1/2$).
The projected operators are $S_{3}^{+}=S_{5}^{+}\rightarrow\frac{2}{\sqrt{8+\Delta_{3}^{2}}}\tilde{S}_{5}^{+}$
and $S_{3}^{z}=S_{5}^{z}\rightarrow\left(1+\frac{\Delta_{3}}{\sqrt{8+\Delta_{3}^{2}}}\right)\tilde{S}_{5}^{z}$,
and, consequently, the effective couplings are 
\[
\tilde{J}_{i}=\frac{2}{\sqrt{8+\Delta_{3}^{2}}}J_{i}\mbox{ and }\tilde{\Delta}_{i}=\left(\frac{\Delta_{3}+\sqrt{8+\Delta_{3}^{2}}}{4}\right)\Delta_{i}.
\]

In the free fermion case $\Delta_{i}=0$, the eigenvalues of $\mathbb{M}$
are $E_{1,2}=\mp\frac{1}{2}\sqrt{J_{3}^{2}+J_{4}^{2}}$ and $E_{3}=0$.
Hence, the doublet is $\left|\tilde{+}\right\rangle =J_{4}\left|++-\right\rangle +2E_{1}\left|+-+\right\rangle +J_{3}\left|-++\right\rangle $
and $\left|\tilde{-}\right\rangle =-J_{4}\left|--+\right\rangle -2E_{1}\left|-+-\right\rangle -J_{3}\left|+--\right\rangle $,
which yields the projections $S_{3}^{+}\rightarrow\frac{J_{4}}{\sqrt{J_{3}^{2}+J_{4}^{2}}}\tilde{S}_{5}^{+}$
and $S_{5}^{+}\rightarrow\frac{J_{3}}{\sqrt{J_{3}^{2}+J_{4}^{2}}}\tilde{S}_{5}^{+}$.
We then arrive at the effective Hamiltonian 
\[
\tilde{H}^{\prime}=\tilde{J}_{2}\left(\frac{S_{2}^{+}\tilde{S}_{5}^{-}+{\rm h.c.}}{2}\right)+\tilde{J}_{5}\left(\frac{\tilde{S}_{5}^{+}S_{6}^{+}+{\rm h.c.}}{2}\right),
\]
with 
\[
\tilde{J}_{2}=\frac{J_{2}J_{4}}{\sqrt{J_{3}^{2}+J_{4}^{2}}}\mbox{ and }\tilde{J}_{5}=\frac{J_{3}J_{5}}{\sqrt{J_{3}^{2}+J_{4}^{2}}}.
\]

Finally, we consider the SU(2)-symmetric case $\Delta_{i}=1$. Here,
the eigenenergies are $E_{1,2}=-\frac{1}{4}\left(J_{3}+J_{4}\pm2\gamma\right)$
(with $\gamma=\sqrt{J_{3}^{2}-J_{3}J_{4}+J_{4}^{2}}$), and $E_{3}=E_{4}=\frac{1}{4}\left(J_{3}+J_{4}\right)$
and the corresponding ground state is $\left|\tilde{+}\right\rangle =\left(-J_{3}+J_{4}+\gamma\right)\left|++-\right\rangle -\left(J_{4}+\gamma\right)\left|+-+\right\rangle +J_{3}\left|-++\right\rangle $.
Thus, the renormalized couplings are 
\[
\tilde{J}_{2}=\left(\frac{2J_{4}-J_{3}+\gamma}{3\gamma}\right)J_{2}\mbox{ and }\tilde{J}_{5}=\left(\frac{2J_{3}-J_{4}+\gamma}{3\gamma}\right)J_{5}.
\]
As expected, $\tilde{\Delta}_{2,5}=\Delta_{2,5}=1$ which preserves
the SU(2) symmetry.

 \bibliographystyle{apsrev4-1}
\bibliography{/home/hoyos/Documents/referencias/referencias}

\end{document}